\documentstyle[aps,multicol,epsf,psfig]{revtex}
\begin{document}
\draft
\title{Numerical model for granular compaction under vertical tapping}
\author{ P.~Philippe, D.~Bideau}
\address{ G.M.C.M., Bat 11A, Campus de Beaulieu,
Universit\'e de Rennes I, F-35042 Rennes, France}
\date{\today} 
\maketitle
\begin{abstract}

A simple numerical model is used to simulate the effect of vertical
taps on a packing of monodisperse hard spheres. Our results are in
good agreement with an experimental work done in Chicago and
with other previous models, especially concerning the dynamics of the
compaction, the influence of the excitation strength on the compaction
efficiency, and some ageing effects. The principal asset of the model
is that it
allows a local analysis of the packings. Vertical and transverse
density profiles are used as well as size and volume distributions of
the pores. An interesting result concerns the appearance of a
vertical gradient in the density profiles during
compaction. Furthermore, the volume
distribution of the pores suggests that the smallest pores, ranging in
size between a tetrahedral and an octahedral site, are not strongly
affected by the tapping process, in contrast to the largest pores
which are more sensitive to the compaction of the packing.

\end{abstract}
\pacs{\bf PACS numbers: 45.05.x, 45.70.Cc, 61.20.-p, 81.05.Rm, 81.40.Cd}
\begin{multicols}{2}

\section{Introduction}
\label{sec:intro generale}

Granular materials constitute the raw materials in a huge number of
human activities as in agriculture, the mining industry, pharmaceutics and
are at the heart of the matter in several ecological concerns as
desertification by eolian erosion or avalanches. Therefore, explaining
a few current granular
processes, such as storage, transport, or collapse, is a real
economical challenge. Furthermore, packings of spheres which is the
simplest model for granular medium, have a great fundamental interest
for physicists: hard sphere systems are indeed a common description
of simple liquids \cite{Bernal}; moreover grains can behave, according to the
external conditions, more or less like a solid, a liquid or a gas
\cite{s/l/g}. This great variety of behaviors for a banal heap of grains makes
granular mechanics a rich area of investigation only
partially clarified at the moment. It is now a well-known result \cite{RLP,RCP1,RCP2}
(although there
is no theoretical explanation for it) that a disordered static packing of equal
hard spheres can cover a large range of volume fraction,
approximately from 56 $\%$, the random loose packing (R.L.P.), 
to 64 $\%$, the random close packing (R.C.P.). For a regular arrangement, the
packing fraction can reach up to 74 $\%$ which corresponds
to the densest structures, namely the hexagonal compact (H.C.)
and the face-centered cubic (F.C.C.) crystals.\\ 
As thermal energy ($k_B$T)
plays no role, because it is insignificant compared to the
gravitational energy of a macroscopic grain, each packing of spheres
is a metastable configuration which can persist as long as there is no
external excitation.
In this frame, the issues of compaction of grains under vertical
taps are a practical way to study the succession of jumps from a
metastable equilibrium to another one. The initial packing is quite
loose and can progressively reach a nearly stationary
configuration (steady state) evaluated through its average volume
fraction. Some experiments done in Chicago \cite{Chicago1,Chicago2,Chicago3} have studied
the influence of the tapping intensity on the steady-state value and the
dynamics of the compaction, which is approximately inverse of the
logarithm of the number of taps. The experimental set-up is a thin
tube of diameter D=1.88 cm filled to about an 80 cm height with
monodisperse, spherical
soda-lime glass beads (of diameter d=1, 2 or 3 mm). The tube is shaken
by an electromagnetic exciter delivering vertical taps, each of them
consisting of an entire cycle of a sine-wave of frequency f = 30
Hz. The excitation strength is parameterized by $\Gamma$, the ratio
between the measured acceleration peak and the gravitational
acceleration g. Moreover, several numerical and theoretical works
\cite{Tetris1,Tetris2,Tetris3,parking1,parking2,Edwards,Head}, most of them dealing with the notions of free volume and
geometric constraint, found the same kind of behavior as obtained
experimentally and some of
them \cite{Tetris2,Tetris3} have pointed out structural ageing effects, as typically
observed in glassy systems. So a parallel might exist between
this granular compaction and the dynamics of out-of-equilibrium systems as
glasses.\\   
In this work, we used a simple model to simulate the compaction of a
packing of monosize spheres submitted to vertical taps. We did not try
to make a realistic description of the quite complex succession of
collisions in a shaken packing: we have kept as the only ingredient
of the model the geometric constraint between hard spheres, which is
believed to be the principal origin of the compaction. Despite the
fact that we
deliberately forgot the mechanical dimension of the problem, the model
is able to reproduce qualitatively the experimental results of the
Chicago group and some further results in agreement with different
numerical and theoretical studies. As the model seems to capture the
physics of the problem, it is then possible to go beyond a global
analysis. Indeed, as a three-dimensional packing of hard spheres,
our description has the quite interesting asset that it is very close to a
real granular medium. So, contrary to almost all the previous works which deal
only with a macroscopic probe (i.e. the average density in all or
part of the packing), our model can provide us
with realistic information on the local structure of a packing and
its evolution under compaction by taps.\\
The paper is organized as follows. A detailed description of the model
is presented in section II. Section III is devoted to the global analysis of
compaction (logarithmic dynamics, hysteresis effect, and ageing
behaviors). In section IV, the local analysis of the packings is
described with the use of
density profiles and size and volume distributions of the pores. Our
conclusions and perspectives end the
paper in section V. 

\section{The model}
\label{sec:model}

The model proposed here is purely geometric and deals only with the steric
constraint, neither friction nor contact law between
the spheres or with the walls is introduced.\\
The different sequences of tapping were initialized from a relatively loose
packing obtained by a steepest-descent algorithm simulating a sequential
gravitational deposition \cite{steepest-descent}. We worked with packings of 4096 spheres of
radius R piled up in a square-box of dimension L = 32R. Concerning
the vertical walls, we used both
periodic boundary conditions (P.B.C.) and fixed boundary conditions
(F.B.C.) i.e. impassable vertical planes. The
top of the box is open whereas the bottom is a fixed impassable plane.\\
A tap
is decomposed in two stages: first a vertical dilation and then a gravitational
redeposition.\\ The first stage corresponds to the external excitation
which will enable the packing to move from a metastable equilibrium
to another one. We used the simplest way to simulate the tap by
applying an uniform dilation $\varepsilon$ to the whole packing (z $\to$
z(1+$\varepsilon$)). This reduction is certainly far from a real tap but we
assume that the way of dilating
the packing is less important than the result of the dilation: a
significant increase of the average free-volume of the spheres which
will allow collective rearrangements during the second stage, the
redeposition of the packing. \\This redeposition procedure must be
non-sequential in order to 
permit such collective behaviors; so we use a monte-carlo algorithm to
discretise the motion of the spheres: a great number of small
displacements are computed . An individual movement procedure is
structured as follows: a sphere, randomly chosen, is submitted to a small
random displacement; if this displacement creates no
interpenetration with another sphere or with the walls (according to 
the boundary conditions), it is accepted otherwise it is
rejected. Because of this binary schema, two neighboring spheres
can not be exactly in contact but, after a sufficient time, they get
very
close to contact. Figure \ref{algorithm}
shows a typical displacement: the values of the polar angle $\phi$ and
of the displacement d
are strictly randomly chosen respectively between 0 and 2$\pi$ and between 0
and $d_{max}$, whereas the choice of the angle $\theta$ follows a random
distribution centered on zero to mimic the effect of gravity. We used
the following Gaussian distribution of width $\theta_0$ truncated
beyond $\pi$/2 in order to orientate all the displacements down to the
bottom of the box.       
\begin{equation}
P(\theta)=A\exp(-(\theta/\theta_{0})^2)
\label{proba-theta}
\end{equation}
The choice of the distribution does not seem to be restrictive: some
attempts with a Poissonian and a linear distribution give qualitatively
the same phenomenology; the pertinent parameter is the width
$\theta_0$. 
\\ With such an algorithm, agitation will persist indefinitely. So we
regularly test the packing during the redeposition process. The
variable checked is $\langle$Z$\rangle$, the average altitude of the packing that is
the average potential energy of the spheres. The redeposition is
stopped when the relative variation of $\langle$Z$\rangle$ becomes
smaller than a threshold $\eta$. The choice of $\langle$Z$\rangle$ is motivated by its easy
evaluation during the process and by its possible link with a statistical
mechanics approach.\\ This simulation is rather close to the one proposed by
Barker and Mehta \cite{simulation1,simulation2} but with some differences
especially concerning the way of introducing gravity and the end of the
redeposition stage.\\ \\
The model uses four parameters: $d_{max}$, $\eta$, $\theta_0$ and
$\varepsilon$. The two first ones have a direct effect on the simulation
time. The smaller $\eta$, the longer the simulation time; still,
$\eta$ must be small enough if we want the redeposition to be nearly
completed. The parameter $d_{max}$ has to be optimised. A very small
value of $d_{max}$ 
allows almost all of the displacements to be accepted but the effect on
the redeposition is very slight and the packing is therefore nearly frozen.
On the contrary, for a large value of $d_{max}$, almost of all the
displacements are refused and, once again, the packing evolves very
slowly. In this study, we used the intermediate
value $d_{max}$ = R/5.\\ $\theta_0$ has a significant effect on the
packing behavior: a very small $\theta_0$ induces a decompaction whereas a
large value decreases the efficiency of the compaction. We found
$\theta_0$ = $\pi$/4 as the optimised
value giving rise to the maximal compaction rate.\\The last 
parameter, $\varepsilon$, corresponds to the external excitation
induced in the packing. This is our control parameter. The value of
$\varepsilon$ can be estimated from experimental results
concerning the dilation of a vertically shaken sand heap \cite{epsilon}:
$\varepsilon$ = $\delta$h/h $\approx$ 5/500 $\approx 10^{-2}$. We can
also try to link roughly $\varepsilon$ to the experimental control parameter,
the dimensionless acceleration $\Gamma$ = A$\omega^2$/g where A and
$\omega$ are, respectively, the amplitude and the frequency imposed to
the bottom of the heap. In first approximation, if we neglect the
loss of energy in the packing, a particle at the top of the heap (z(0) = h)
acquires an initial speed $\omega$A and achieves
 a ballistic flight. Its maximal altitude is $z(0^+)=h
 +\frac{g}{2}(\frac{\Gamma}{\omega})^2$ and then it comes:
\begin{equation}
\Gamma = \omega(\frac{2h}{g})^{1/2}\varepsilon^{1/2}\ 
\Rightarrow\  \Gamma \propto \varepsilon^{1/2}
\label{control parameter}
\end{equation}
As $\varepsilon^{1/2}$ is linked
to $\Gamma$, we will use it as our control parameter to quantify the strength
of the tapping process.\\With this, it is possible
to compare the results of our model with the experimental work of the
Chicago group and with other numerical and theoretical models, almost
all of them dealing only
with a global description of the granular system.  
 
\section{Global analysis}
\label{sec:global}

This global analysis is achieved with different
average values. We did not use a direct evaluation of the packing fraction from
the number of spheres in a reference volume because wether boundary
effects are significant or, for a smaller
volume, the statistics become
too poor. Moreover, the choice of the reference volume is not unique:
it can be, for example, the space that contains all the spheres or the
smaller one that contains only the centers
of the spheres. To avoid being partial, we
evaluate the
packing fraction by averaging the surface packing fraction, $\Phi$, 
calculated on many horizontal cuts. This measure is permissible
because of the following stereologic result: the average
surface fraction of any cut in a packing is equal to the
volume fraction of the packing \cite{Oger}; with horizontal cuts, this calculus
is just a spatial integration which gives the exact volume fraction. 
The quantity $\Phi_{b}$ is
calculated in this way at the bottom of the packing between the heights 0 and
4R; $\langle\Phi^{c}\rangle$ comes from a similar calculation on
approximately 90 $\%$ of the packing and is corrected near the bottom wall
by a perturbated zone model \cite{Ben Aim}. This model uses a corrective
factor for the average density of a packing near a wall (between 0 and R) with
regard of a packing not perturbated by any wall. For the case of
spheres near a plane, this factor is estimated to 16/11.
It is also interesting to study $\langle$Z$\rangle$, the average
potential energy of the whole system, which is quite pertinent in a
statistical mechanics description.    

\subsection{The dynamics of compaction}
\label{sec:dynamics}

The densification of the packing is observed through the temporal
evolution of the preceding mean values; here the time is the number of
taps and what we call dynamics of the compaction is, in fact, the
succession of metastable equilibrium, each jump from one to another being
induced by the taps.
Figure \ref{fig-dynamics} shows compaction laws obtained with fixed
boundaries (F.B.C.) and three different excitation rates. This excitation
intensity $\varepsilon^{1/2}$ has a decided effect on the compaction
dynamics (see section \ref{sec:hysteresis}). The simulation curves are in
good agreement with the experimental data and compatible with the
following fit previously proposed \cite{Chicago1}:
\begin{equation}
X(t) =X_{\infty} -
\frac{\Delta X_{\infty}}{1+B_{X}\ln(1+{t/\tau_{X}})}
\label{fit-log X}
\end{equation}
with X = $\Phi_{b}$ or $\langle\Phi^{c}\rangle$. For $\langle$Z$\rangle$, a
nearly similar fit can be proposed:
\begin{equation}
\langle\rm{Z}\rangle(t) = 
\langle\rm{Z}\rangle_{\infty} 
\Bigg(
\frac
{1+B_Z\ln(1+{t/\tau_Z})}{\frac{\langle\rm{Z}\rangle_\infty}
{\langle\rm{Z}\rangle_0}+B_Z\ln(1+{t/\tau_Z})} 
\Bigg)
\label{fit-log Z}
\end{equation}
We have noticed that a sum of two exponentials can also fit
$\langle$Z$\rangle$(t) reasonably well.\\
The dependance of these parameters on $\varepsilon$ is
difficult to characterize. We simply
note that the parameter B is consistent with an exponential dependance on
$\varepsilon^{1/2}$ (i.e. $\Gamma$). \\
This compaction dynamics is quite particular: as the packing
progressively densifies, the compaction efficiency decreases. So the
dynamics reduces speed and the system evolves to a steady-state
without never really reaching it. This slowing down is particularly
remarkable for the smallest
values of $\varepsilon^{1/2}$. This specific dynamics requires the
study of the densification on a logarithmic time scale.\\
It is also interesting to analyse the fluctuations of the curves,
especially when the packing becomes close to its asymptotic
or steady-state (SS) limit. The power spectrum of the fluctuations
X-X$_{SS}$ as a function of the frequency, i.e. the inverse of the
taps number, shows more or less a simple power-law in a
log-log diagram (with a slope in the range 1 to 1.5). The effect of $\varepsilon$ is noticable only for the
high frequencies. Moreover, the simple standard deviation of the fluctuations,
$\sigma_X$ = $\sqrt{\langle\rm{(X-X}_{SS})^2\rangle}$, seems to
be directly proportional to $\varepsilon^{1/2}$ or $\Gamma$. These results,
calculated for $X=\langle$Z$\rangle$, are
presented in figure \ref{fig-fluctuations}.\\
Furthermore, we have noticed that the periodic boundary conditions do
not qualitatively affect these observations; the same remark can be made
concerning the results of the following section.   

\subsection{Hysteresis on the steady-state values}
\label{sec:hysteresis}

The next stage consists on studying the influence of the excitation
parameter $\varepsilon^{1/2}$ on the maximal value of the packing
fraction. For this purpose, we carried out a succession of simulations
with a sequence of 4000 taps. The steady-state value is estimated by
averaging the packing fraction on the 1000 last taps or directly
through  the last value. The smaller is
$\varepsilon^{1/2}$ and the larger is the difference between this steady-state 
value and the asymptotic value $X_{\infty}$ given by the fit. Moreover, in this
small excitations range,
$X_{\infty}$ can get over the R.C.P. limit and $\tau$, the characteristic 
time of the fit, increases spectacularly. In fact, the fit becomes more and 
more uncertain in so far as the triplet ($X_{\infty}$, B, $\tau$) is no 
longer unique, and depends strongly on the range of taps over which the data 
fitting is performed. This deviation
between the steady-state value and an uncertain asymptotic limit has also 
been noticed in
the experimental work of the Chicago group \cite{Chicago3} and in some theoretical 
studies \cite{parking2,Head}. The dependance of
$\langle\Phi^{c}\rangle$ on $\varepsilon^{1/2}$ is shown in figure
\ref{fig-hyst} 
(solid black squares). The different packings are obtained after 4000 taps 
of strength $\varepsilon$, starting from the same initial packing. The
 curve has a bell shape with a maximum between 0.1 and 0.2.\\
\\
If we now compute a unique tapping sequence with a progressive
increase of the excitation $\varepsilon^{1/2}$ after every 4000 taps 
(constant excitation increment: $\Delta\varepsilon^{1/2}=+0.025$),
we obtain nearly the same curve for $\langle\Phi^{c}\rangle$, as 
can be seen in figure \ref{fig-hyst} (open circles).
When carrying out the same process in the opposite way 
i.e. with a progressive decrease of $\varepsilon^{1/2}$ 
($\Delta\varepsilon^{1/2}=-0.025$), 
two things can happen:\\
If, while increasing, $\varepsilon^{1/2}$ went beyond a
critical value of  $(\varepsilon^{1/2})^{\star}$ $\approx 0.15$, the
final packing fraction $\langle\Phi^{c}\rangle$
does not decrease but increases a bit more to a maximum value. 
If we compute 
another increase process ($\Delta\varepsilon^{1/2}=+0.025$), we cover
approximately the same values. This last
upper branch, including the part above $(\varepsilon^{1/2})^{\star}$ is 
represented on figure \ref{fig-hyst} (up and down open triangles). As it is
relatively well reproductible, it is called ``reversible''. We can also 
notice on this reversible branch that $\langle\Phi^{c}\rangle$ decreases 
with $\varepsilon^{1/2}$.\\
On the contrary, if $\varepsilon^{1/2}$ stayed below
$(\varepsilon^{1/2})^{\star}$ during the increase stage, the
steady-state values do not evolve significantly; they are nearly frozen, 
and it is hard to estimate whether
there is a compaction or a decompaction process,because the dynamics is very
slow. This last branch is called ``irreversible'' and reflects the
great metastability of the corresponding packings.\\
To summarize, there is a strong hysteresis effect
 which allows the maximum
compaction rate to be reached by an $\varepsilon^{1/2}$ increase-decrease
sequence. These observations are in very good agreement with the results of
Nowak $\&$ al \cite{Chicago2}. In particular, Figure \ref{fig-hyst} is to be compared to the
experimental data
obtained with 1mm diameter beads, corresponding to an aspect-ratio of nearly
19, close to that used in our simulation (L/2R = 16). Surprisingly, for an aspect-ratio
of 9, the experimental results show a much larger increase of the
packing fraction on the reversible branch, up to nearly 66 $\%$
(i.e. more than the R.C.P. limit which may indicate a commensurability
between the cylinder and the beads \cite{Chicago2}). However, for a still
smaller aspect-ratio of 6, the reversible branch is  more similar to
the first case, with a moderate increase to a maximal value below the
64 $\%$ limit.

\subsection{Ageing}
\label{sec:ageing}

In these kind of systems in slow evolution to a final equilibrium, it
is possible to demonstrate ageing effects by comparing the system at different
ages. This comparison can be made by use of temporal correlation functions
of global values ($\rho$,$\langle$Z$\rangle$, ...) between the initial
packing and the same packing after an evolution time $t_W$ (waiting
time). In this study we work with the following function: 
\begin{equation}
A(t,t_W) =
\overline{(\langle\rm{Z}\rangle(t)-\langle\rm{Z}\rangle(t+t_W))^2}
\label{ageing functions}
\end{equation}
Here, $\overline x$ indicates the statistical average of x;
that is, the mean value calculated for a certain number of realizations
of the same experiment. The results have been averaged
on only 10 realizations because of the limitation due to the calculation
time. The statistics are, therefore, rather poor, that is why we use solely
$\langle$Z$\rangle$, which fluctuates quite less than the other global
values. In figure \ref{fig-ageing}
are drawn the curves of A(t,t$_W$) obtained for different values of
t$_W$. There is obviously a scaling law; a similar fit as in
section \ref{sec:dynamics}, with the three parameters $A_\infty$
(the asymptotic limit), $B_A$ and $\tau_A$, is quite compatible 
with the data:
\begin{equation}
A(t,t_W) = A_\infty\bigg(1-\frac{1}{1+B_A ln(1+t / \tau_A)}\bigg)
\label{ageing fit}
\end{equation} 
The same kind of ageing effects have already been pointed out in previous
numerical studies \cite{Tetris2,Tetris3}. These effects
confirms the great similarity between granular compaction or more generally slow
granular rheology and glassy systems submitted to time-dependent
driving forces (see for instance \cite{Glass1,Glass2,Glass3}).
\\
\\
To conclude with the global analysis of the compaction, it is
satisfying to note that our simulation reproduces qualitatively well the
previous results obtained both experimentally and theoretically. This model
seems to capture most of the physics of the problem. Because it gives a
very realistic description of a granular system as a three-dimensional
packing of hard spheres, it can be a quite useful and interesting
tool to go beyond a global description to a local analysis
of the packings' structure during the compaction process. 

\section{Local analysis}
\label{sec:local}

To study locally the packings of spheres more or less compacted, we
use two kind of descriptions: packing fraction (or density)
profiles are calculated vertically and transversly to the
box, and size and volume distributions of the pores in a packing are
evaluated and then analysed.

\subsection{Density profiles}
\label{sec:profiles}

Using the surface packing fraction calculated by stereological
cuts (as in the evaluation of $\Phi_b$ and $\langle\Phi^{c}\rangle$), we can have access
to vertical (horizontal cuts) and transverse
(vertical cuts) density profiles.\\
Some examples of vertical profiles are shown in figure \ref{vert-prof}. These
have been obtained with fixed boundary conditions (F.B.C.), but the use
of periodic boundaries (P.B.C.) induces no significant differences.
The profiles are
characterised in particular by a negative vertical gradient $\alpha$ and 
by large peaks near the bottom of the box. These peaks reflect a
partially ordered packing due to the wall and are very close to previous
experimental observations \cite{Benenati}. The gradient can be roughly estimated in an
intermediate zone (5 $\le$ z/R $\le$ 22 for F.B.C. and 5 $\le$ z/R
$\le$ 26 for P.B.C.) after
smoothing the profile. This gradient, directly linked to
$\varepsilon$, is qualitatively different from previous numerical results
\cite{Tetris3}, where a local densification is obtained at the interface. It could
be objected that this gradient comes
directly from the modeling of the tap through an uniform
dilation. Nevertheless, despite the fact that $\alpha$ is difficult to estimate very
precisely, it does not seem to be monotonic with $\varepsilon^{1/2}$,
but has more or less the same kind of bell-shaped dependance as the other
steady-state values. This behavior can not be caused only by the
dilation. But, in contrast to the other average values, it seems that
$\alpha$ presents no hysteresis effect which denotes a relatively
different behavior. In conclusion, the origin of the anisotropy of the
packing, observed through this gradient $\alpha$, is not well
understood. It may come from both the uniform dilation of the system 
and its specific redeposition under the simulated particle motion
under gravity.\\
Being inspired by a Fermi level profile \cite{Fermi}, we
can propose the following average fit for a typical vertical profile:

\begin{equation}
\Phi(\overline{z}) = 
\frac
{\Phi_0 -\alpha \overline{z}}
{1+exp(\beta(\overline{z}-\overline{z}^*))}
\quad \hbox{where} \quad \overline{z}=z/R 
\label{vertical profile fit}
\end{equation}

Figure \ref{trans-prof} presents a few transverse profiles in fixed
boundaries; they are qualitatively close to experimental profiles
\cite{Vanel}. Here again, some peaks indicate a local organisation in
layers due to the walls; this effect has approximately a three
layers range. The
average lateral density increase (at a
distance less than 7R from the walls corresponding roughly to this wall
effects range) is noted $\delta\Phi_{lateral}$
as is the central increase $\delta\Phi_{central}$. The last one is
systematically smaller than the other. Both of them are calculated in
comparison with the initial profiles and reflect the spatial
repartition of the bulk compaction. These profiles with periodic
boundaries reveal no peak, due to the absence of walls. The central
zone is a bit larger but keeps the same qualitative shape and
densifies as well during a tapping sequence. This
observation of an obvious compaction even in P.B.C. ensures that
compaction is not, or at least not principally, due to wall effects.
This was not evident considering the small aspect ratio used
in the experience of the Chicago group.
Quantitatively, the absolute value of
the packing fraction is larger in the periodic conditions but its increase
due to compaction is a bit smaller.
\\As global values,
$\delta\Phi_{lateral}$ and $\delta\Phi_{central}$ have the same
dependence on $\varepsilon^{1/2}$ (bell shaped
curves) than the others. 
It is also
possible to study their evolution with the number of taps. The results,
presented in figure \ref{increases}, point out, once again, the nearly frozen
dynamics for small values of $\varepsilon^{1/2}$.\\
Moreover, we can remark that the initial packing in
F.B.C. presents a great metastability. This one is particularly
noticeable on the transverse profile (see fig. \ref{trans-prof}) with an
``under-population'' of the spheres near the
walls. This explains the significant compaction of
$\delta\Phi_{lateral}$ caused
by the first tap (see fig. \ref{increases}). This metastability is due to the
construction of the initial packing: it was built by a gravitational
algorithm [16] with periodic conditions, and a slight agitation was then
induced in the packing to adapt it to fixed boundary conditions (F.B.C.). 
This last stage was not sufficiently efficient. 
        
\subsection{Size and volume distributions of the pores}
\label{sec:pores}

Another way to analyse a packing of particules is to study the
interstitial voids. This void-space is more difficult to apprehend
because, in contrast to a particle, a cavity has no geometric
limit. We then introduce the notion of pore as the ``void'' between four
neighboring spheres. Previous studies have already been made on this
issue, both theoretically and experimentally. Gotoh \cite{gotoh} introduced the pore
size distribution $P_0$ as the probability for a randomly positioned sphere
of radius r' to intercept no particle center. He proposed a theoretical
expression for $P_0$ derived from the Percus-Yevick approximation
which agrees well with previous results on random close packings
\cite{Finney,Tory}:

\begin{eqnarray}
0\le\sigma\le1,
\quad P_0(\sigma) = &&1-\Phi\sigma^3\\
1\le\sigma,
\quad P_0(\sigma) = &&
(1-\Phi)exp[\frac{\Phi}{(1-\Phi)^2}(-(1-2\Phi)\sigma^3\nonumber
\\&&+9/2\Phi\sigma^2+1-5/2\Phi)]
\label{gotoh distribution}
\end{eqnarray}
\\
where $\sigma$ is the ratio r'/R (R is the radius of the hard spheres) 
and $\Phi$ is the average volume fraction. Figure \ref{fig-gotoh}
confronts this expression with the distribution calculated in one of
the packing obtained after a compaction sequence of 4000 taps ($\Phi =60.6
\%$) and with periodic boundaries. The distribution calculated in the initial 
packing ($\Phi = 58.4 \%$) is also
represented and is quite close to the other. This slight
difference means that the distribution $P_0$ is insufficiently sensitive to
small structural changes as compaction.\\   
\\
Hence we find that it is more efficient to work with a direct
statistical analysis on the size of the
pores. To do this, we use the {Vorono\"\i} tesselation of a packing
\cite{Voronoi}: a Vorono\"i polyhedon around a sphere is the region
of space in which all the points are closer to this given sphere than
to any other. Two neighbors correspond to two {Vorono\"\i}
polyhedra that share a face. Each vertex is equidistant from 
the center of four neighboring
spheres and therefore constitutes a pore. More precisely, 
we define the pore as the virtual sphere in contact
with these four neighboring spheres which interpenetrates none of them. The
size of the pore is then the radius of this ``void-
sphere''. The volume of this sphere reflects partially the total
void-volume situated inside the tetrahedron formed by the centers of the
four neighboring spheres. In a packing, it is possible to calculate the
size distribution of the pores $\rho_{\xi}(\xi)$, where $\xi = r/R$ with r the
radius of a pore and R the radius of the hard spheres. The
normalisation of the distribution gives $\int
\rho_{\xi}(\xi)d\xi=N_{P} $, 
the total number
of pores in the measurement volume. This distribution is linked to
Gotoh's one. Thus, $P_0(\sigma=r'/R)$ is more or less the sum
of the pores of size greater than r'. Therefore $P_0$ is a cumulative
distribution in comparison with $\rho_{\xi}$, which is expected to be
more sensitive to the local structure. A rather similar analysis by
use of the size distribution of the pores was
previously sketched out by Barker and Mehta \cite{simulation1}.\\ 
If we now use directly the normalized volume $v=\xi^3=\frac{4/3\pi
  r^3}{4/3\pi R^3}$
as new variable, the corresponding statistical density is
$\rho_{v}(v)=\frac{\xi}{3}\rho_{\xi}(\xi)$ 
(here the ``volume'' of a pore is reduced
to $4/3\pi r^3$). The distribution
$v\rho_v(v)$ reflects the contribution of the pores to the total
porosity according to their size; this last distribution seems to be the more pertinent in
problems of free volume and compaction. We have noted that, by integrating $v\rho_v(v)$, a new global
value is obtained and corresponds to the average normalized volume of a
spherical pore: $\nu=\frac{1}{N_P}\int v\rho_v(v)dv=\langle{v}\rangle/V$, where
$\langle{v}\rangle$ is the average
volume of a pore, V is the volume of the hard spheres, and $N_P$ is the number
of pores. The average pore volume $\nu$ has the
same dynamics and the same kind of reversible-irreversible behavior
as described in section \ref{sec:hysteresis}. 
Figure \ref{fig-dist} a) shows the distributions $v\rho_v(v)$ for a
given packing at different
stages of its compaction (in F.B.C. and with an excitation strength
$\varepsilon=2 \times 10^{-2}$). 
The statistics are calculated in a
smaller box of height 18R located at a distance 2R from the
walls to avoid some boundary problems. In this measurement volume, $N_P$
varies approximately from 9550 to 10250 for the different packings analysed.
These different volume distributions $v\rho_v(v)$ are slightly
affected by the taps in the small pore domain, whereas the variation 
of the packing
fraction clearly appears in the progressive reduction of the tail of
the distribution in the large
pore zone. So, there is more or less a persistence of the distribution for the
values of v approximately limited by the volume of an octohedral
site ($\xi_O=\sqrt{2}-1 \approx 0.414$ and $v_O \approx
0.0711$). The distribution $v\rho_v(v)$ is bell-shaped with an 
over-population for
the largest pores (the tail) and with a minimum size of the pores
corresponding to a tetrahedral site ( $\xi_T=\sqrt{3/2}-1 \approx
0.225$ and $v_T \approx 0.0114$). With respect to the small pores, this is 
in contradiction with a Poisson distribution
proposed in a previous theorical model for logarithmic dynamics
\cite{Boutreux}. But, in figure \ref{fig-dist} b), we note that,
in  the range of volume corresponding to the tail of the distribution, 
$\rho_{v}(v)$ is compatible with a Poisson law: 
$\rho_{v}(v)\propto e^{(-v/v_0)}$, where $v_0$ is directly linked to 
$\nu$, the average normalized volume already defined, or to 
$\langle\Phi^{c}\rangle$, the average packing fraction.\\   
\\These results must be compared with previous work on the
issue. First, Bernal \cite{Bernal}  analyzed the arrangements of
spheres by characterizing the cavities between the spheres. To do
this, he studied the different polyhedra formed by the sphere centers
as corner. He found five canonical holes. Table \ref{tab1} \cite{Frost} presents
his results obtained on a mechanical model of hard spheres and concerning
the statistical weight (in number and in volume) and the $\xi$-value
corresponding to each hole. In fact, these canonical holes are more or
less distorted (otherwise $\rho_{\xi}(\xi)$ would be an addition of Dirac
peaks) and the $\xi$-value corresponding to the regular hole
is therefore a lower limit. So we can note that the
tetrahedron and the tetragonal dodecahedron correspond to the smallest
values of $\xi$ for which we have shown that the volume distribution
$v\rho_v(v)$ is not greatly affected by the tapping process. In
contrast, the octahedron, the trigonal
prism and the archimedian antiprism (the two last in very low
proportions) appear in the large pores range and are consequently more
sensitive to the compaction state of the packing. So, according to
this Bernal's classification on the structure of the pores, we can
suggest that compaction is principally due to rearrangements of the three
largest canonical pores.  

\section{Conclusions and perspectives}
\label{sec:conclusion}

A simple model of hard spheres under vertical taps based solely on
geometric constraint is sufficient to
qualitatively describe previous experimental and numerical results:
the same
kind of compaction dynamics, hysteresis effect on the steady-state
values, and ageing
behaviors. The originality of this model, i.e. a realistic description
of a granular system as a three-dimensional packing of hard spheres, permits a
structural analysis of the packings. A semi-local
study on density profiles suggests the existence of a negative
vertical gradient in the packings but with no clear hysteresis
effect. It also confirms a compaction in the
bulk which can not be caused only by wall effects, which are
particularly noticeable with fixed boundaries (F.B.C.). A more local
analysis on the void space of the packings shows, in a model of spherical
pores, a volume distribution sensitive to the packing fraction for the
large pores and nearly stationary for the small ones. Compaction could then
be principally explained by collective rearrangements of the largest pores.\\  
Further to this numerical work, an experimental study of the compaction
induced by vertical taps is being carried out. The packing fraction is
deduced from a measure of absorption of a horizontal gamma-rays
beam. In addition to the average volume fraction in the
bulk, our set-up permits the  evaluation of the vertical density profile in
the packing; these measurements would be crucial in order to test the
results of our numerical model, especially
concerning the existence of a negative vertical gradient. Furthermore,
compaction is studied in quite different experimental conditions
than in the previous work of the Chicago group. In this later one, the
set-up is a tube of height 80 cm and with an approximate diameter
of 2 cm filled with 1 mm diameter soda lime glass beads. So, transverse
wall effects are very significant and the
vertical pressure on the packing is saturated for almost all of the height
of the heap, the overload being completely held up by the walls. Inversely, the cylinder used in our set-up has a
diameter of 10 cm and a height of 15 cm; about 80 percent of it is
filed with 1 mm diameter glass beads. Here, the wall effects become
negligible and the vertical pressure is definitely not saturated
in the packing. It will be interesting to see to which extent these
differences can qualitatively and quantitatively affect the 
compaction under vertical tapping.      

\section{Acknowledgements}
\label{sec:thanks}  

We are very grateful to
R. Jullien for his support in numerical questions especially
concerning his gravitational deposition algorithm. We also wish to
thank P. Richard who has constructed the program of size
distribution of the pores and J. Jenkins who nicely read this
manuscript and ran a spelling check of it.

\end{multicols}

%figures

\begin{figure}[ht]
\caption{A typical displacement during the redeposition stage
  of the algorithm.}
\label{algorithm}
\end{figure}

\begin{figure}[ht]
\caption{Bottom packing fraction $\Phi_b$ versus
t, the number of taps, in logarithmic (up) and linear (down) time scale
for three excitation rates ($\varepsilon  = 5 \times 10^{-3}$,
$5 \times 10^{-2}$, and $1.5 \times 10^{-1}$). 
The solid lines are the simulation results and
the dotted lines are the inverse-logarithmic fits.}
\label{fig-dynamics}
\end{figure}

\begin{figure}[ht]
\caption{The power spectrum of the fluctuations of 
$\langle$Z$\rangle$ versus frequency (the inverse of 
the number of taps) (up) and the simple standard deviation of 
$\langle$Z$\rangle$, $\sigma_{\langle\hbox{Z}\rangle}$, which is nearly 
linear with $\varepsilon^{1/2}$ (down).}
\label{fig-fluctuations}
\end{figure}

\begin{figure}[ht]
\caption{Steady-state values of
  $\langle\Phi^{c}\rangle$ obtained after 4000 taps with
different values of $\varepsilon^{1/2}$ (solid black squares) 
and  hysteresis during
a sequence of increase (open circles), decrease (open up-triangles)
and increase again (open down-triangles) of the excitation with an
increment $\Delta\varepsilon^{1/2}$ every 4000 taps.}
\label{fig-hyst}
\end{figure}

\begin{figure}[ht]
\caption{Ageing effects on the time-correlation function
  A(t,t$_W$) for several waiting times t$_W$ : the different curves (up)
  and a collapse according to the fit in dotted line (down).}
\label{fig-ageing}
\end{figure}

\begin{figure}[ht]
\caption{Two examples of vertical density profiles with
  their fit : for the initial packing (dotted line) and for a packing
  obtained after 4000 taps with $\varepsilon = 10^{-1}$ (solid line).}
\label{vert-prof}
\end{figure}

\begin{figure}[ht]
\caption{Two examples of transverse density profiles :
  for the initial packing (dotted line) and for a packing
  obtained after 4000 taps with $\varepsilon = 10^{-1}$ (solid line). The
  two vertical dotted lines indicate the frontiers in the calculus of
  $\delta\Phi_{lateral}$ and $\delta\Phi_{central}$.}
\label{trans-prof}
\end{figure}

\begin{figure}[ht]
\caption{Lateral (up) and central (down) packing fraction
  increases (in comparison with the initial packing) versus t, the
  number of taps for 4 values of $\varepsilon$.}
\label{increases}
\end{figure}

\begin{figure}[ht]
\caption{Theorical pore size distribution $P_0$ (thick
  solid line) compared with the numerical calculation for a packing with the
  same average volume fraction (open circles). No significant difference with
  the calculations for the initial packing (thin solid line) which has
  a quite smaller volume fraction.}
\label{fig-gotoh}
\end{figure}

\begin{figure}[ht]
\caption{a) Volume distribution of the
  pores $v\rho_{v}(v)$ for a
  packing at different stages of its compaction (F.B.C. and
  $\varepsilon=2 \times 10^{-2}$): t+1 = 1, 100, and 10000.
 As the packing progressively densifies (i.e. as t increases),
  the tail of the distribution corresponding to the largest pores
  tends to vanish (as symbolized by the arrow). Here, T and O
  indicate the v-values for tetrahedral and
  octahedral sites in a dense packing (F.C.C. or H.C.).
  b) Plot of $ln(\rho_v(v))$ versus v for the same packings. The
  different tails are compatible with a Poisson law.}
\label{fig-dist}
\end{figure}

%table

\begin{table}[p]
\begin{tabular}{|c||c|c|c|}
& $\xi$=r/R & Number ($\%$) & Volume ($\%$) \\
\hline
tetrahedron & 0.225 & 73.0 & 48.4 \\
\hline
half-octahedron & 0.414 & 20.3 & 26.9 \\
\hline
trigonal prism & 0.528 & 3.2 & 7.8 \\
\hline
tetragonal dodecahedron & 0.353 & 3.1 & 14.8 \\
\hline
archimedian antiprism & 0.645 & 0.4 & 2.1 \\
\end{tabular}
\caption{Characteristics of the Bernal canonical holes.} 
\label{tab1}
\end{table}

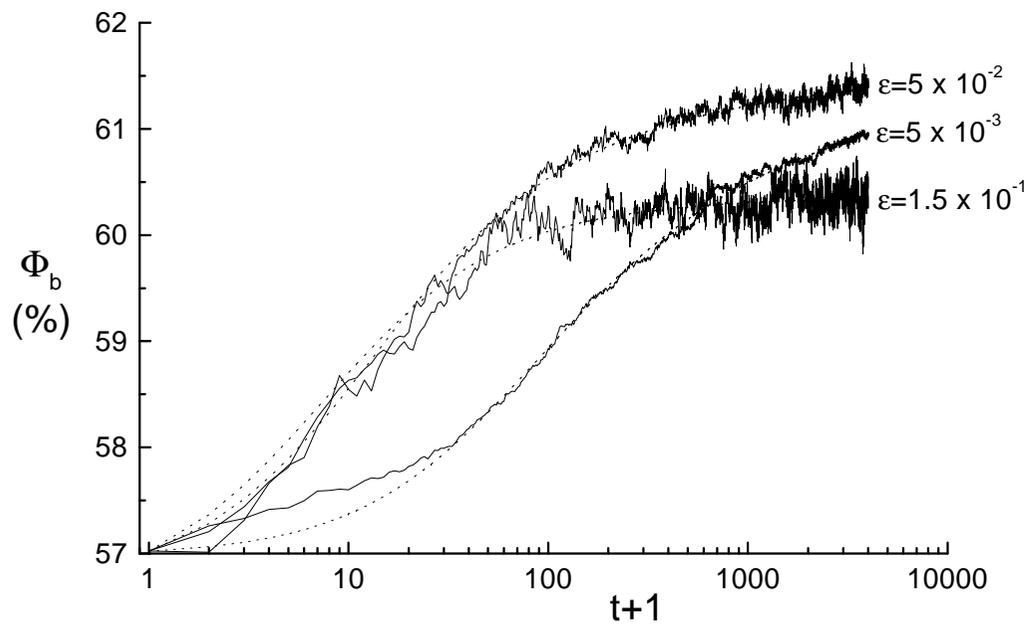
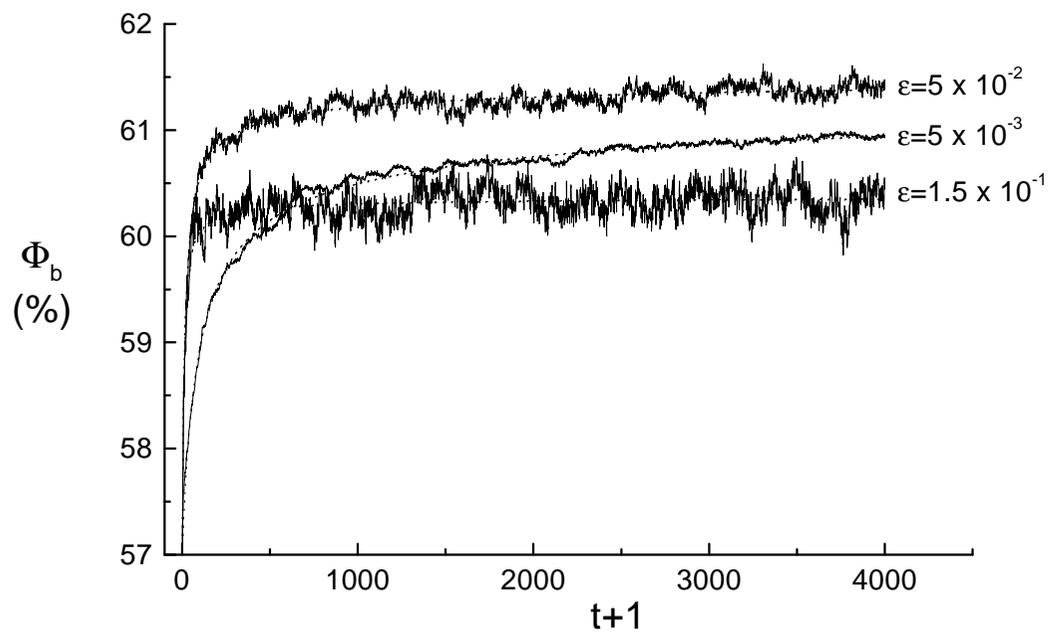
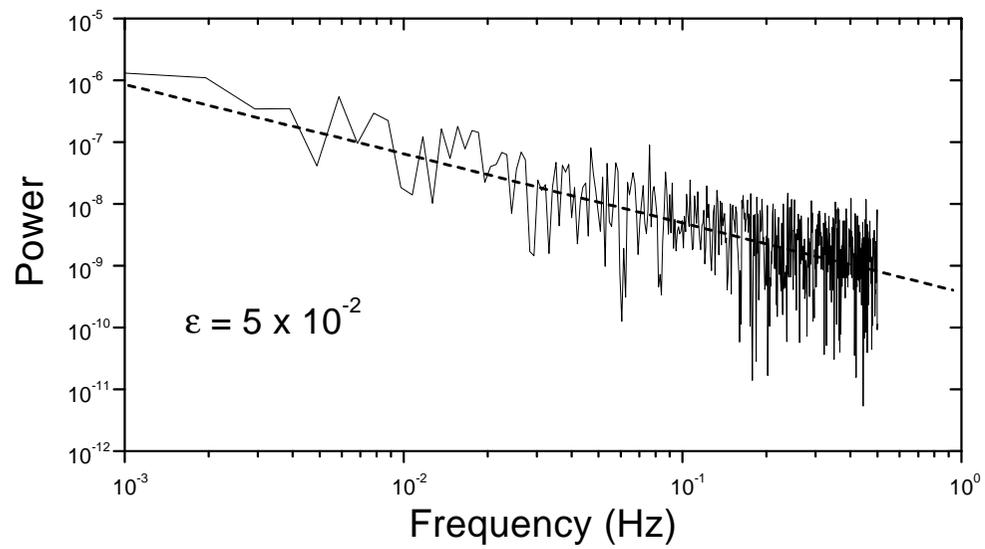
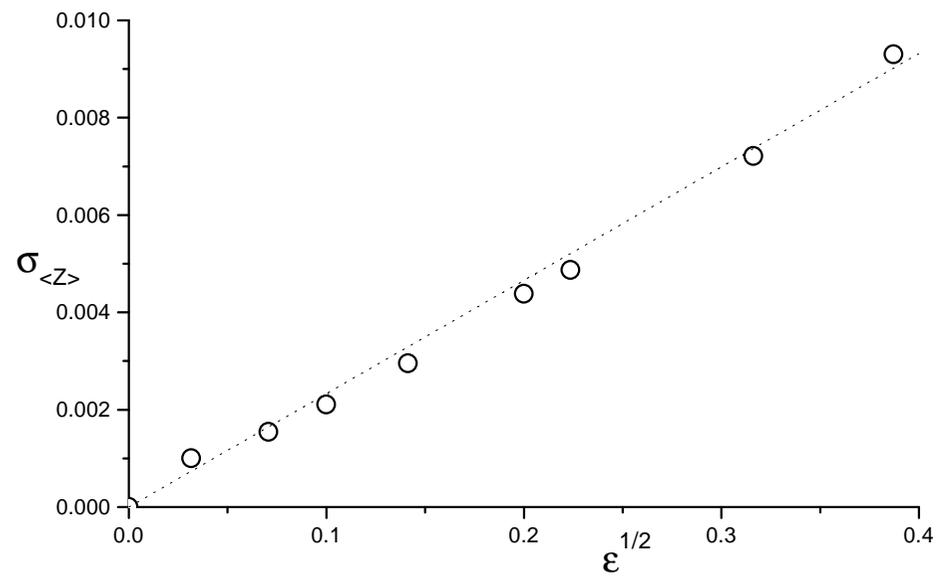
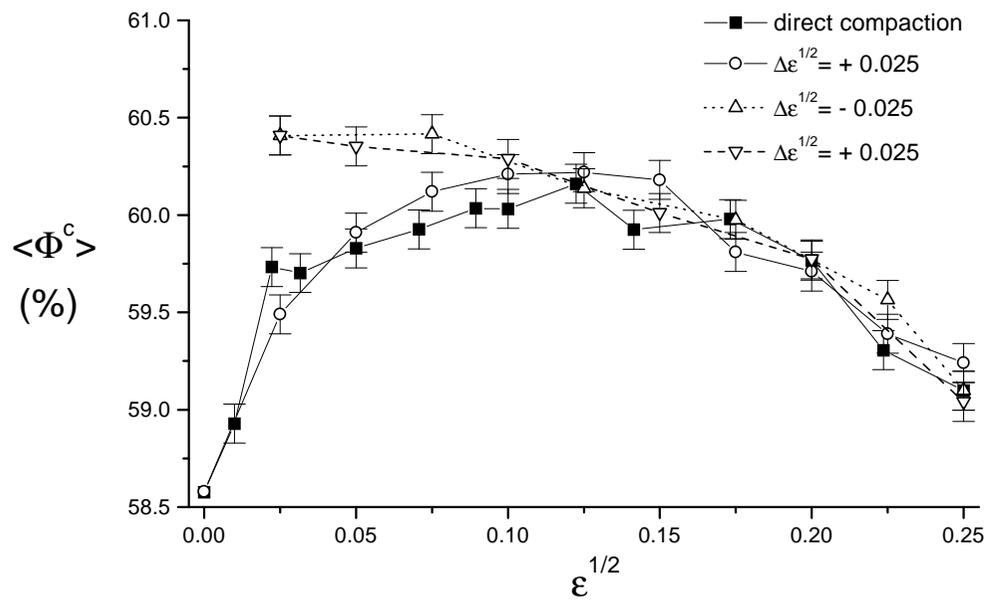
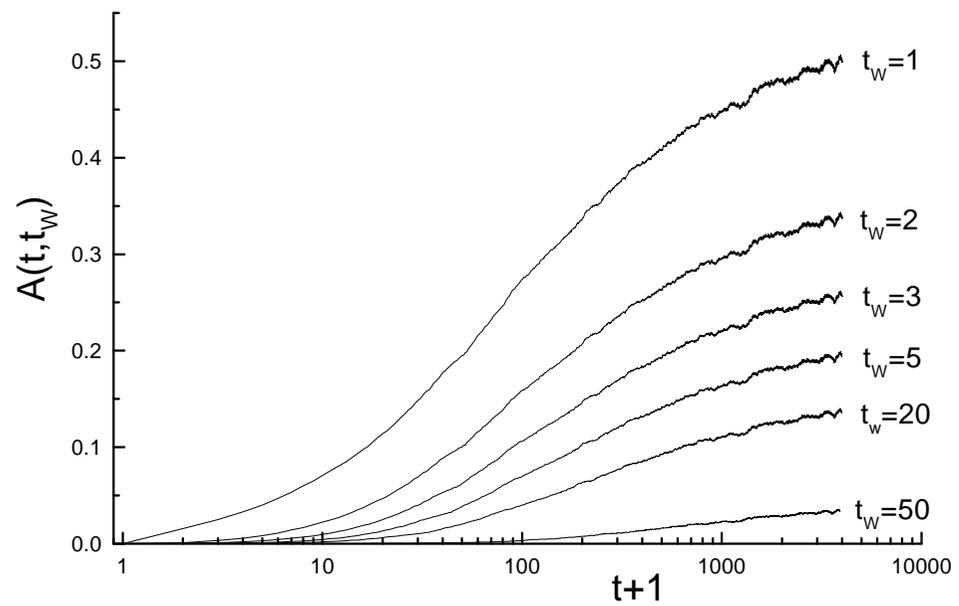
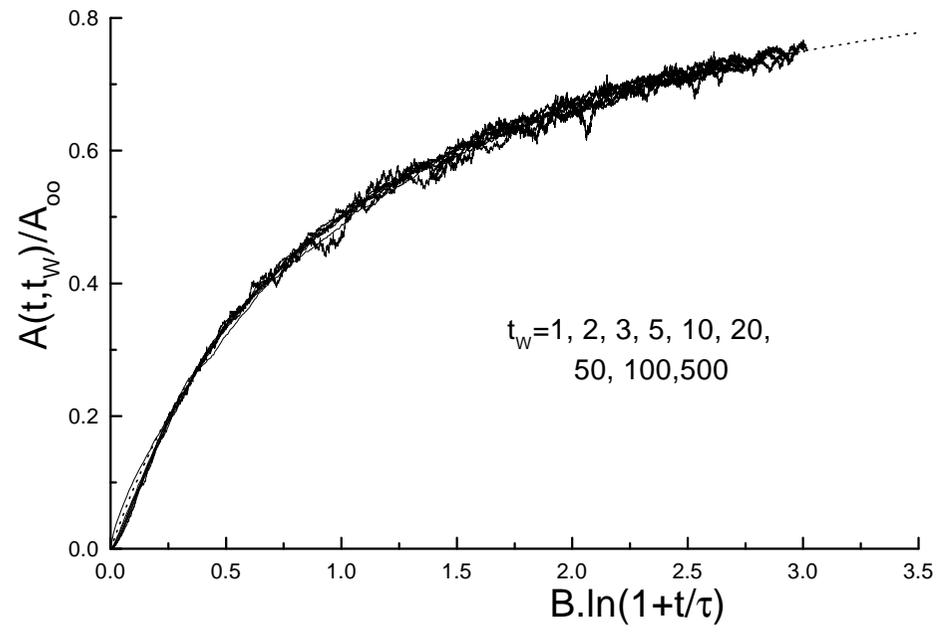
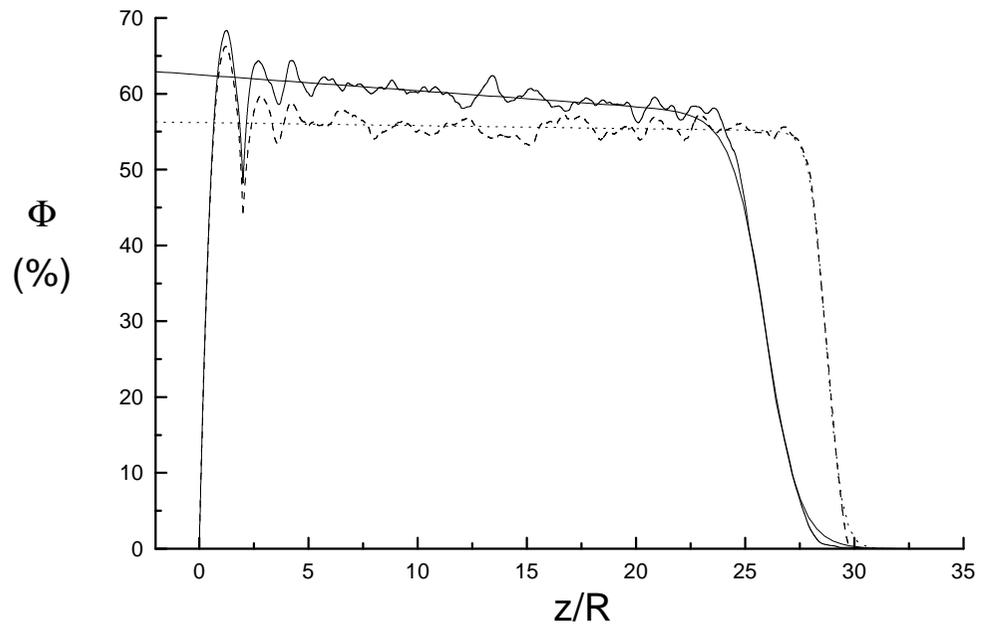
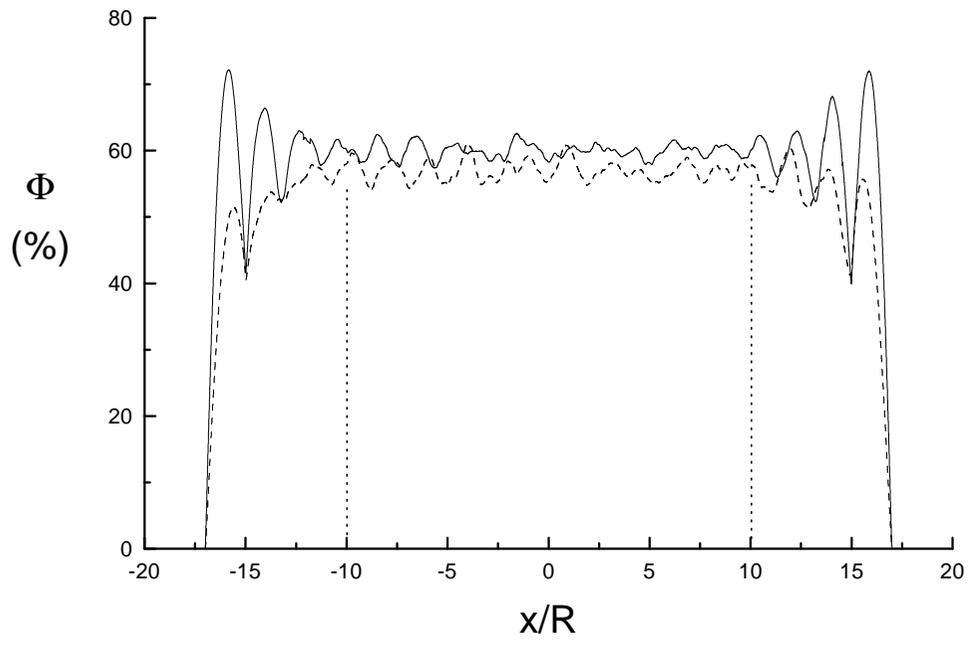
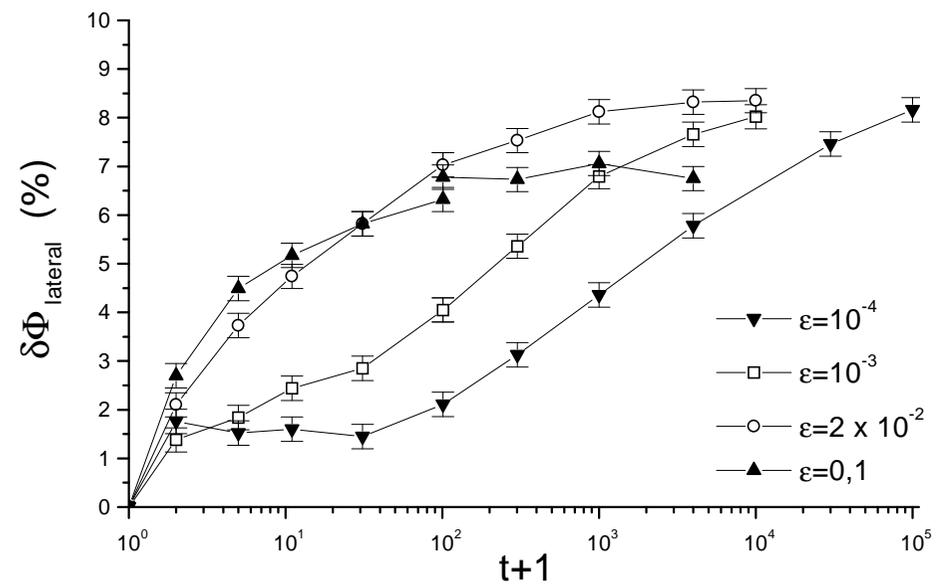
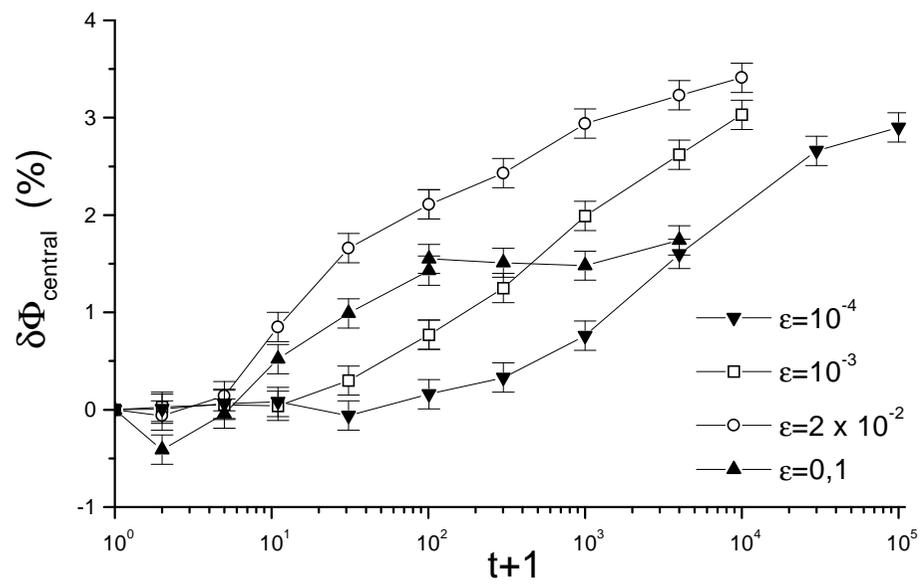
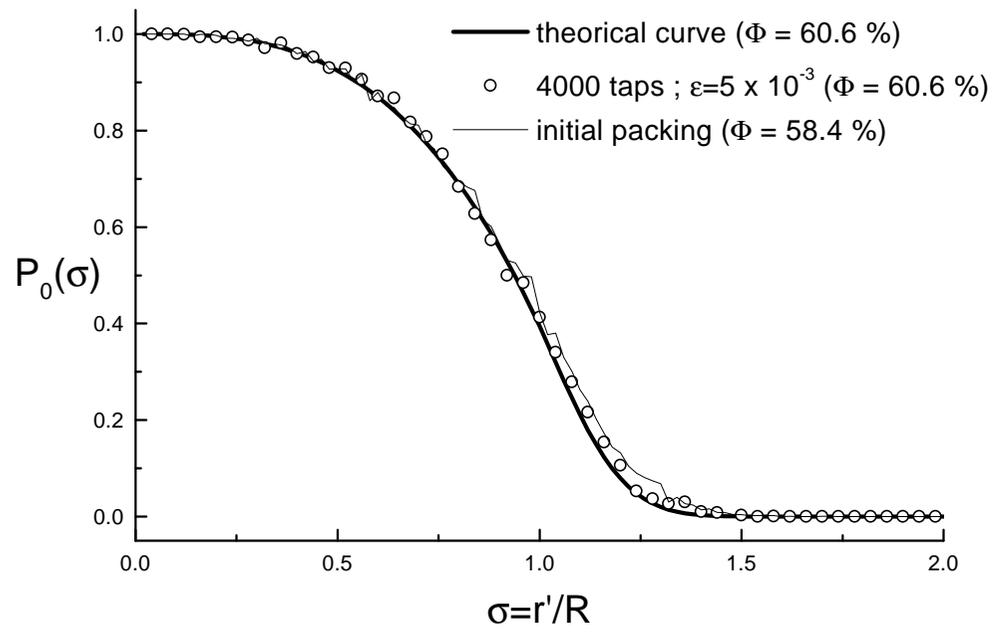
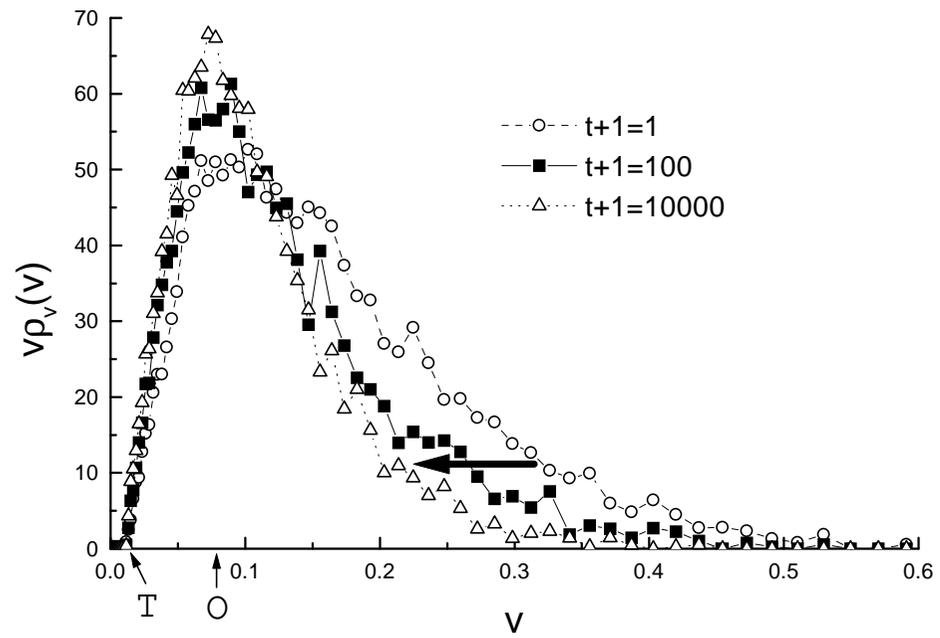
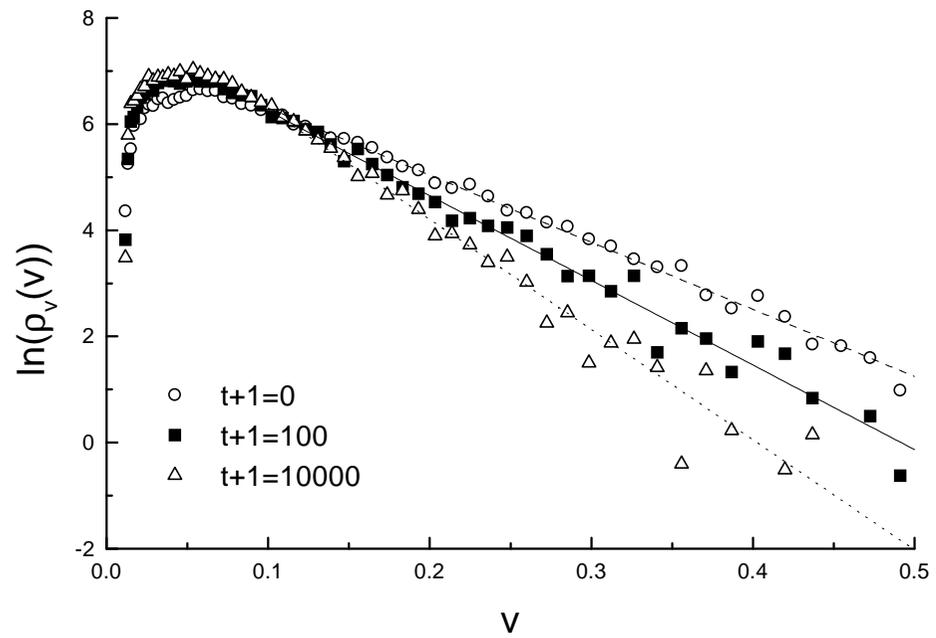
\end{document}